\documentstyle[12pt,fleqn,espcrc1]{article}

\newcommand{\AmS}{{\protect\the\textfont2
  A\kern-.1667em\lower.5ex\hbox{M}\kern-.125emS}}
\newcommand{\be}{\begin{equation}}
\newcommand{\ee}{\end{equation}}
\newcommand{\ba}{\begin{eqnarray}}
\newcommand{\ea}{\end{eqnarray}}
\newcommand{\bann}{\begin{eqnarray*}}
\newcommand{\eann}{\end{eqnarray*}}
\newcommand{\ul}{\underline}

\newcommand{\bit}{\begin{itemize}}
\newcommand{\eit}{\end{itemize}}
\newcommand{\ben}{\begin{enumerate}}
\newcommand{\een}{\end{enumerate}}
\newcommand{\bc}{\begin{center}}
\newcommand{\ec}{\end{center}}
\begin{document}
\hbadness=10000
\title{Hydrodynamical Beam Jets in High Energy
Hadronic Collisions\footnote{to appear in the Proceeding of
Quark Matter 93} }
\author{\underline{U. Ornik}
\address{ GSI Darmstadt, Darmstadt, F.R.Germany}
\thanks{E. Mail: ORNIK@TPRI6B.GSI.DE},
R. M. Weiner
\address{ Physics Department, University of Marburg,
F.R.Germany}
\thanks{E. Mail: WEINER@VAX.HRZ.UNI-MARBURG.DE} and
G.Wilk
\address{ SINS, Nuclear Theory Department, Warsaw, Poland}
\thanks{E. Mail: WILK@FUW.EDU.PL}
}
\maketitle

\begin{abstract}
A study of hadronic data up to TEVATRON energies in
terms of relativistic hydrodynamics indicates an extended 1-dimensional stage
of the expansion which suggests a  jet like
behaviour of the fireball along the collision axis.
\end{abstract}

The Landau Hydrodynamical Model (LHM) exists already for 40 years but there
has been little progress in understanding its successes. Hydrodynamics
assumes in general local thermal equilibrium (l.e.), a condition difficult
to realize in  small and short lived hadronic systems with a typical size
of $10^{-13}$ cm and a corresponding lifetime of $\sim  10^{-23}$ sec.
The discovery of subentities of hadrons (quarks and gluons) with the
associated proliferation of degrees of freedom has facilitated the believe
in l.e. \cite{VH}, nevertheless the phenomenological success of the LHM
has not been understood so far. The situation is better for heavy ion
reactions which are larger systems, and where it is easier to get
l.e.

We want to suggest that one possible reason for the success of
LHM  in \ul{hadronic} reactions is the fact that Landau \cite{landau}
and most of his followers  used  {\em only} a 1-dimensional ($1d$)
solution which corrects for the possible absence
of l.e.
in these reactions.
This conclusion follows from a comparison of $1d$ and $3d$ solutions to be
 reported below.
The $1d$ approach assumes
that the strongly
compressed initial fireball expands at first mainly in the longitudinal
direction (the width of the rapidity distribution is directly connected with
the strength of this flow).
Then a
conical ($3d$) correction follows.
A reasonable estimation for the moment $\tau_{3d}$ when the conical
expansion starts
($R$
denotes the transverse radius of the fireball) is:
\be
\tau_{3d}=\sqrt{t^2-x^2}=a_{3d}R \label{eq:tau3d}
\ee
where $a_{3d}$ is a phenomenological parameter (determined by Landau from
simple geo\-me\-tri\-cal considerations \cite{landau} to be equal to
$a_{3d}\equiv a_L=(1+c_0^2)/c_0^2$;  $c_0$ is the speed of sound).

In Fig. 1 we present rapidity distributions calculated for
``conical'' $1d$
and $3d$ \cite{OPW}
solutions compared with $\bar{p}p$ data at SPS ($\sqrt s=20$ GeV) and
ISR ($\sqrt s=53$ GeV) energies.
The 3d solution uses $c_0$ as given by lattice QCD \cite{OPW} and
$K=$0.35 at $\sqrt{s}=20$ GeV and $K=0.176$
at $\sqrt{s}=53$ GeV. More realistic values of $K$ lead even to a
worsening of the agreement with data. The 1d solution uses $c_0^2=0.18$
at $\sqrt{s}$ = 20 GeV and $c_0^2=0.195$
at $\sqrt{s}=53$ GeV and $K=0.5$.

It turns out that
only the $1d$ solution is able to fit the data with reasonable values
for $c_0$ and inelasticity $K$.
The transverse
expansion (present only in the 3d case) develops at the expense
of the
longitudinal expansion and therefore reduces the width of the
rapidity distribution.
However, going to still higher energies we have found
that one has to increase the duration of the $1d$ stage even
further (by increasing
$a_{3d}$ above the limit $a_L$ given by Landau).
In Fig. 2 we show fits to different pseudorapidity distributions which are
obtained with $a_{3d}$ increasing from $\sqrt2a_L$ at $\sqrt{s}=53$ GeV to
$\sqrt{10}a_L$ at $1800$ GeV. We have checked that these results hold
(almost) independently of the  concrete variant of initial conditions and
equation of state (EOS) provided they are physically reasonable.

This observation  poses a serious problem
 for LHM because the
corresponding extended $1d$ stage is not present in the ``real'' $3d$
dynamics (which is based on the assumption that each fluid cell has in its
rest frame an isotropic pressure - a result of the assumed isotropic momentum
distribution corresponding to local equilibrium (l.e.)). One might therefore
 argue that
 because of the breakdown of l.e. conventional hydrodynamics
  is not valid anymore. On the other hand, the success
of $1d$ LHM in  describing data,
illustrated above,
allows also a different in\-ter\-pre\-ta\-tion.
In the following we shall argue that at high energies a ``new" physical
effect occurs, namely
a strong anisotropy in the flow caused by  some physical pro\-ces\-ses acting
on top of the conventional hydrodynamical description.

The simplest ``model" for such an anisotropy would be to postulate the
exis\-ten\-ce
of beam jets associated e.g. with the leading particles. This would make
necessary a reformulation of the in\-elas\-ti\-ci\-ty effect in the LHM. So far
inelasticity $K$ was considered in the LHM by assuming that only the function
$K$ of the available energy contributed to the mass of the fireball which
underwent hydrodynamical expansion \cite{OPW,carr}. In this way the leading
particles ``had done their job" and did not interfere anymore with the central
fireball. This treatment may be an oversimplification.
A more realistic approach in this direction
is represented by the two component
model proposed in \cite{fow}. It is based on
the  analysis of
multiplicity distributions $P(n)$
at energies
between 20 and 540 GeV \cite{ua5}.
They were interpreted as indicating
the presence of two different types of sources
emitting secondaries: $(i)$ - {\it chaotic},
provided by gluonic interaction (i.e., equilibrated) and
concentrated in the central rapidity
(i.e., hydrodynamical)  region with
$P(n)$ of negative binomial type and $(ii)$ - {\it coherent},
provided by the leading valence quarks (therefore far from equilibrium)
and extending over the entire rapidity region (but contributing mainly
to fragmentation region) with $P(n)$ consistent with
a Poisson distribution. In this context
the anisotropy and the elongation of $\tau_{3d}$   can be
viewed as a manifestation of the {\it coherent} component due to
the leading valence quarks\footnote{
Another, more radical
possibility would be to assume that the configuration space in high energy
hadronic reactions has fractal nature meaning effectively that the number
of dimensions $d$ is less than 3.}.

Formally these possibilities could be formulated in terms of anisotropic
hy\-dro\-dy\-na\-mics as proposed in \cite{greiner}. In the present study
 we shall limit
ourselves just to consider the extended $1d$ stage as a phenomenological
observation which may have
 consequences for the
interpretation of data
from future experiments
for hadronic and heavy ion collisions (RHIC, LHC or SSC). Here
the hadronic reactions provide a lower
limit for stopping, lifetime and
equilibration and an upper limit for the width of the rapidity
distribution $\sigma$. For this last quantity
we get as an upper limit
\be
\sigma\le \sqrt{\ln R/ \delta_i}+\sigma_{therm}; \qquad
(\sigma_{therm} < 1.65~~~{\rm for}~~~T_f < 0.2~{\rm GeV})\label{sigma}.
\ee
where $\delta_i$ is the initial longitudinal extension of the fireball and
$\sigma_{therm}$ the contribution of the thermal motion to the rapidity
width.
It grows with energy slower than the phase space.
This is shown in Fig. 3.
A hydrodynamic
stage in high energy collisions leads therefore
to a limited value of $\sigma$
(e.g. $\sigma$ $<$ 5 at LHC energies).
It can be also
shown that all relevant information concerning the transition from a
strongly interacting non-equilibrium system to a thermalized fireball
is contained in the fragmentation region (i.e., the phase space region
where the transition from a local equilibrium in  a pre-equilibrium
stage takes
place.
This is also illustrated in Fig.4 where one can see how the initial
longitudinal
size of the fireball (determined in the pre-equilibrium stage)
is strongly reflected in the shape of rapidity distribution. The main effects
appear
in the fragmentation region.

We conclude than that a hydrodynamical
analysis of $\bar pp$ data
indicates
a large
extension of the 1d stage of the expansion
and
is described approximately
by the Khalatnikov solution \cite{landau}. The observation of the
fragmentation region ($3<y_{cm}<5$) is
essential for the investigation of
the transition from the
pre-equlibrium to the  local equilibrium stage of the reaction.

This work was supported in part by the Deutsche Forschungsgemeinschaft,
the Gesellschaft
f\"ur Schwerionenforschung and the Polish State Committee for Scientific
 Research,
Grant  No. 2 0957 91 01.

\end{document}